\documentclass[aps,prl,reprint,superscriptaddress,letterpaper,showpacs]{revtex4-1}
\usepackage[english]{babel}
\usepackage{ucs}
\usepackage[utf8x]{inputenc}
\usepackage{amsmath}
\usepackage{amsfonts}
\usepackage{amssymb}
\usepackage{graphicx}
\usepackage{bm}
\usepackage{bbm}
\usepackage{empheq}
\usepackage{color}

\newcommand{\bra}[1]{\langle #1|}
\newcommand{\ket}[1]{|#1\rangle}

\newcommand{\mm}[1]{\mathrm{#1}}
\newcommand{\ui}{\mathrm{i}}
\newcommand{\ue}{\mathrm{e}}

\newcommand{\abs}[1]{\left|#1\right|}

\begin{document}

\title{Coherent Adiabatic Spin Control in the Presence of \\ Charge Noise Using Tailored Pulses}
\author{Hugo Ribeiro}
\author{Guido Burkard}
\affiliation{Department of Physics, University of Konstanz, D-78457 Konstanz, Germany}
\author{J. R. Petta}
\affiliation{Department of Physics, Princeton University, Princeton, New Jersey 08544, USA}
\affiliation{Princeton Institute for the Science and Technology of Materials (PRISM), Princeton University, Princeton,
New Jersey 08544, USA}
\author{H. Lu}
\author{A. C. Gossard}
\affiliation{Materials Department, University of California at Santa Barbara, Santa Barbara, California 93106, USA}

\date{\today}

\begin{abstract}
We study finite-time Landau-Zener transitions at a singlet-triplet level crossing in a
GaAs double quantum dot, both experimentally and theoretically. Sweeps across the
anticrossing in the high driving speed limit result in oscillations with a small
visibility. Here we demonstrate how to increase the oscillation visibility while keeping
sweep times shorter than $T_2^*$ using a tailored pulse with a detuning dependent level
velocity. Our results show an improvement of a factor $\sim 2.9$ for the oscillation
visibility.  In particular, we were able to obtain a visibility of $\sim 0.5$ for
Stückelberg oscillations, which demonstrates the creation of an equally weighted
superposition of the qubit states.
\end{abstract}

\pacs{73.23.Hk, 72.25.-b, 73.21.La, 85.35.Gv}

\maketitle

The adiabatic theorem of quantum mechanics states that a quantum system will remain in
its instantaneous eigenstate if the variation of a dynamical parameter is slow enough on
a scale determined by the energy separation from other eigenstates~\cite{born1928}.
However, there are systems for which adiabaticity breaks down resulting in a transition
between states.  The first result quantifying population change in such a process is due
to independent works by Landau, Zener, Stückelberg, and
Majorana~\cite{landau1932,zener1932,stuckelberg1932,majorana1932}.  They considered a
coupled two-level quantum system whose energies are controlled by a time dependent
external parameter, which is defined such that the system exhibits an anticrossing of
magnitude $\Delta = 2 \lambda$ at $t=0$. If the system is prepared in its ground state,
$\ket{0}$, at $t = -\infty$ and swept through the anticrossing by modifying the external
parameter in such a way that the energy difference is a linear function of time, $\Delta
E = \alpha t$, then the probability to remain in $\ket{0}$ at $t=\infty$ (in the diabatic
basis) is given by $P_{\mm{LZSM}} = \ue^{-\frac{2 \pi \lambda^2}{\hbar \alpha}}$, which
is known as the Landau-Zener(-Stückelberg-Majorana) (LZSM) nonadiabatic transition
probability.  Remarkably, this elegant solution, although valid only in the asymptotic
limit for an infinitely long sweep, has demonstrated its accuracy in real physical
systems for which the sweep has a finite duration~\cite{shevchenko2010}.

Another success of the asymptotic formulation resides in an accurate description of LZSM
interferometry. If the system is driven back and forth across an anticrossing, it
accumulates a Stückelberg phase that gives rise to periodic variations in the transition
probability~\cite{shevchenko2010}. Although the exact accumulated phase can only be
calculated by solving the time-dependent Schrödinger
equation~\cite{ribeiro2009,ribeiro2010,sarkka2011,fuchs2011,studenikin2012}, a scattering approach
assimilating the phase acquired in a single passage to a Stokes phase~\cite{meyer1989}
nicely reproduces experimental results obtained in superconducting
qubits~\cite{oliver2005}, two-electron spin qubits at a singlet ($S$)-triplet ($T_+$)
anticrossing ~\cite{petta2010,gaudreau2012}, and in nitrogen-vacancy centers in
diamond~\cite{huang2011}.

\begin{figure}[b]
\includegraphics[width=1\columnwidth]{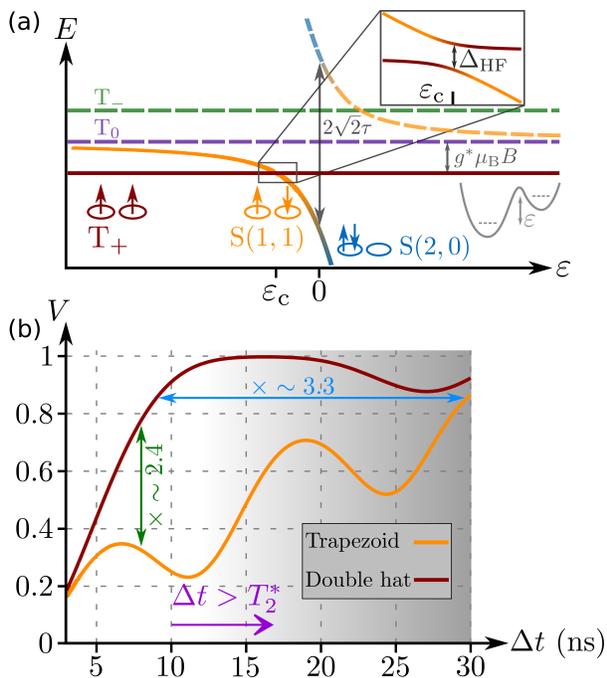}
\caption{(color online) (a) DQD energy levels as a function of the detuning,
$\varepsilon$, near the $(1,1) \leftrightarrow (2,0)$ charge transition. The low energy
hybridized singlet state and the triplet $\mm{T}_+$ form a qubit whose dynamics can be
controlled through LZSM interferometry by sweeping the system through the hyperfine
mediated anticrossing. (b) Comparison of Stückelberg oscillation visibility $V$ as a
function of pulse length, $\Delta t$, for a trapezoid and ``double hat'' pulse with same
maximal amplitude. The oscillation visibility is calculated within a finite-time LZSM
model, where it is given by $V=4 P_{\mm{LZSM}}(1-P_{\mm{LZSM}})$. ``Double hat'' pulses
allow for more than a factor of 2 improvement while keeping $\Delta t < T_2^*$.
}
\label{fig:energydiagram}
\end{figure}

Focusing on spin qubits, passage through a $S$-$T_+$ anticrossing in the energy level
diagram is analogous to a spin-dependent beam splitter~\cite{petta2010}. There are two
major challenges relating to quantum control of such systems. First, in two-electron
double quantum dots (DQD), the $S$-$T_+$ anticrossing is located near the $(1,1)
\leftrightarrow (2,0)$ interdot charge transition, where $(N_{\mm{L}}, N_{\mm{R}})$ refer
to the number of electrons in the left and right quantum dots. As a result, the singlet
state involved in the spin-dependent anticrossing is a superposition of $(1,1)$ and
$(2,0)$ singlet states. Second, the magnitude of the splitting at the level anticrossing
is set by transverse hyperfine fields. To achieve LZSM oscillations with 100\%
visibility, the sweep through the anticrossing would have to be performed on a timescale
set by the electron spin decoherence time $T_{\mm{2}}^*$.  As a result, there is a
tradeoff between adiabaticity and inhomogeneous dephasing. While there are several
studies about dissipative adiabatic passages (see, for instance,
\cite{ao1989,gefen1991,wubs2006,saito2007,nalbach2009,lehur2010}), it remains to be shown
how to make a system less sensitive to dissipation while at the same time increasing
adiabaticity.

\begin{figure*}[t]
\includegraphics[width=2\columnwidth]{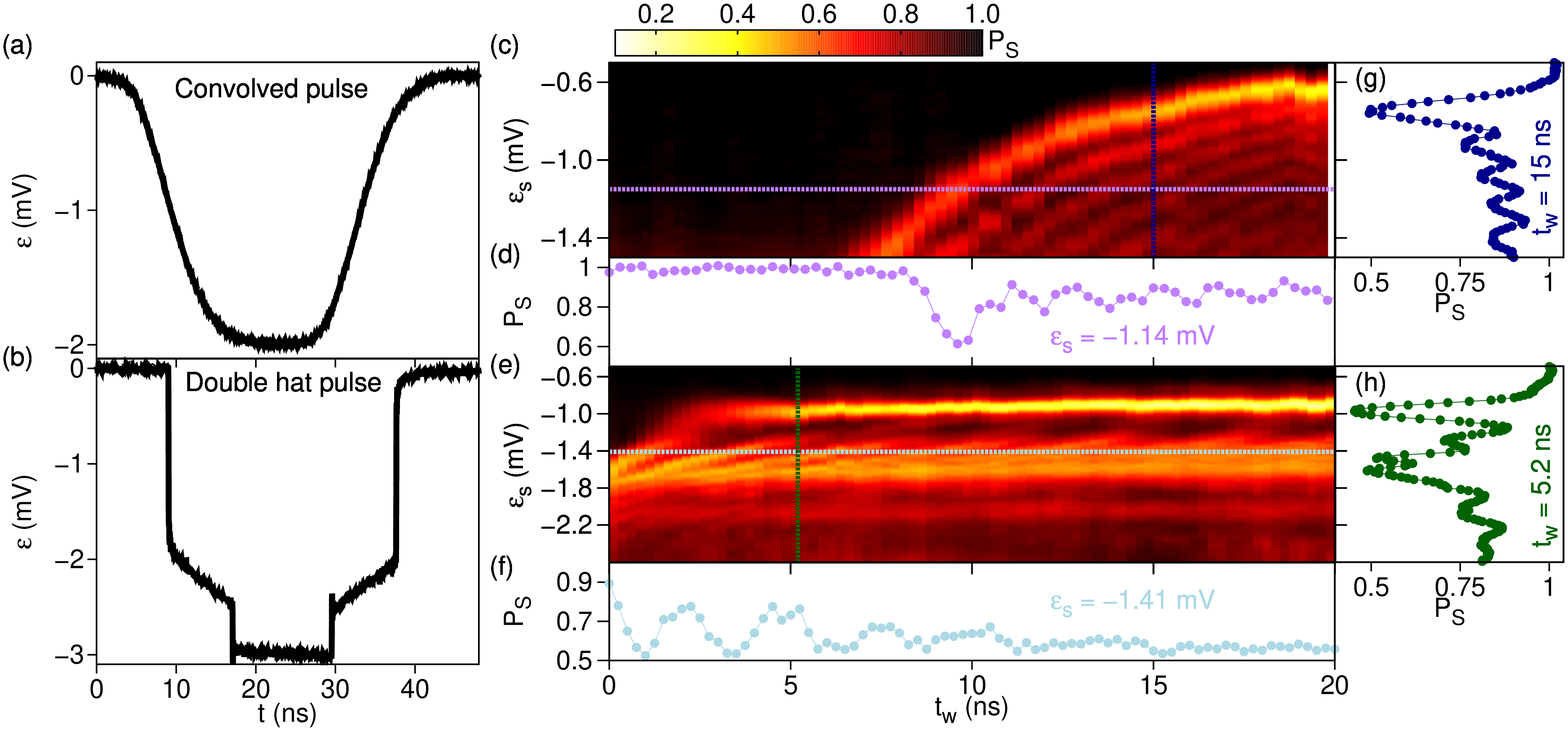}
\caption{(color online) (a) Convolved pulse obtained by convolving a trapezoid pulse
with a Gaussian pulse. (b) ``Double hat'' pulse with a detuning-dependent level velocity.
(c) Singlet return probability $P_{S}$ as a function of $\varepsilon_{\mm{s}}$ and
$t_{\mm{w}}$ for $B=50\,\mm{mT}$ using convolved pulses.  (d) Trace taken along
$\varepsilon_{\mm{s}}=-1.14\,\mm{mV}$. (e) Singlet return probability $P_{S}$
measured with ``double hat'' pulses  plotted as a function of $\varepsilon_{\mm{s}}$ and
$t_{\mm{w}}$ for $B=55\,\mm{mT}$. The results exhibit a high-visibility region
corresponding to slow level velocity portion of the ``double hat'' pulse. (f) Trace taken
along $\varepsilon_{\mm{s}}=-1.41\,\mm{mV}$. This value of $\varepsilon_{\mm{s}}$
corresponds to a passage through the anticrossing with the slow level velocity portion
of the pulse. (g) and (h) Traces taken along $t_{\mm{w}}=15\,\mm{ns}$ and
$t_{\mm{w}}=5.2\,\mm{ns}$ for convolved and ``double hat'' pulses respectively. A
comparison between the different traces shows that the ``double hat'' pulse allows us to
achieve higher visibilities, while keeping the total pulse duration below the limit set
by $T_2^{*}$.}
\label{fig:expresults}
\end{figure*}

In this Letter, we attempt to reconcile the contradiction between the need for a slow
(adiabatic) passage susceptible to dissipation and a fast passage minimizing dissipation
effects. Our approach is based on the observation that the biggest population change
occurs in the vicinity of the anticrossing. We have developed a multi-ramp pulse
sequence that has a detuning dependent level velocity, which we refer to as ``double
hat'' pulse [see Fig.~\ref{fig:expresults}(b)]. The slow level velocity portion of the
pulse is chosen to coincide with the passage through the $S$-$T_+$ anticrossing in
order to increase the visibility of the quantum oscillations.

To demonstrate the advantages of ``double hat'' pulses, we consider a finite-time LZSM
model~\cite{vitanov1996}. In this model, there are three parameters that control the
magnitude of $P_{\mm{LZSM}}$: the dimensionless coupling $\eta = \lambda /\sqrt{\alpha
\hbar}$ and the dimensionless initial and final times $T_{\mm{i,f}}=
\sqrt{\frac{\alpha}{\hbar}}t_{\mm{i,f}}$, where $t_{\mm{i,f}}$ are the start and stop
times for the pulse relative to $t=0$ defined at the anticrossing. The dependence on
$T_{\mm{i,f}}$ results in oscillations of $P_{\mm{LZSM}}$. In
Fig.~\ref{fig:energydiagram}(b), we plot the visibility of Stückelberg oscillations,
given by $V=4 P_{\mm{LZSM}}(1-P_{\mm{LZSM}})$ \cite{shevchenko2010}, as a function of pulse duration for
trapezoid (single ramp) and ``double hat'' pulses. The duration of the pulse is increased
by lowering the level-velocity $\alpha$. For the ``double hat'' pulse, only the slow
level velocity is changed. To be consistent with the regime studied in experiments, we
choose $\lambda = 50\,\mm{neV}$ and energy differences on the order of the Zeeman
splitting ($-\Delta E_{\mm{i}} = \Delta E_{\mm{f}} = 2.5\,\mu\mm{eV}$).  The results
demonstrate that ``double hat'' pulses can improve the oscillation visibility while maintaining a short
pulse duration. The oscillation visibility is enhanced because ``double hat'' pulses
allow a passage through the anticrossing with a slower level velocity $\alpha$ as
compare to trapezoid pulses.  The oscillatory behavior of the results are a consequence
of the finite-time LZSM model.

Ideally, one would like the visibility to be unity, which corresponds to the perfect beam
splitter limit, $P_{\mm{LZSM}} = 0.5$. Its achievement would imply the possibility of
realizing the Hadamard gate, which is essential to perform certain quantum algorithms
(e.g.\ Shor's period finding algorithm~\cite{shor1997}). Optimization methods to obtain
high-fidelity adiabatic passages (i.e. $P_{\mm{LZSM}}=0$) have already been
studied~\cite{torosov2011}.

We measure and model LZSM transitions at the $S$-$T_+$ anticrossing for finite
duration sweeps. Measurements are performed on a GaAs/AlGaAs heterostructure that
supports a two-dimensional electron gas located $110\,\mm{nm}$ below the surface of the
wafer. We use a triple quantum dot depletion gate pattern, where two of the dots are
configured in series as a DQD and the third dot serves as a highly sensitive quantum
point contact charge detector~\cite{petta2010,som}. The DQD is
configured in the two-electron regime, where the electrons can either be separated in the
$(1,1)$ configuration or localized on a single quantum dot, forming the $(2,0)$ charge
state. In this regime, the spin states are the singlets $S(2,0)$ and $S(1,1)$
and the (1,1) triplet states $T_+$, $T_0$, and $T_-$. Interdot tunnel
coupling $\tau$ results in hybridization of the charge states at zero detuning with a
resulting splitting of magnitude $2 \sqrt{2} \tau$ between a ground and excited state
singlet, that we respectively denote $S$ and $S'$. An external magnetic field
is applied perpendicular to the sample, resulting in Zeeman splitting of the triplet
states, as depicted in Fig.~\ref{fig:energydiagram}(a). The hyperfine interaction
between electron and nuclear spins results in an anticrossing between $S$ and
$T_+$ located at $\varepsilon_{\mm{c}}$. The energy difference at the anticrossing,
$\Delta_{\rm HF}$, is set by transverse hyperfine fields~\cite{taylor2007}.

Simulated interference patterns are obtained by solving the master equation $\dot{\rho} =
-\frac{\ui}{\hbar}\left[H,\rho \right] + \frac{1}{2}\sum_{i=1}^{3} \left( \left[L_i \rho
, L_i^{\dag}\right] + \left[L_i, \rho L_i^{\dag}\right]\right)$~\cite{lindblad1976}.
Here, the Hamiltonian $H$ describes the dynamics in the vicinity of the $S$-$T_+$
anticrossing and is given by~\cite{ribeiro2011},
\begin{equation}
	H (t)=E_{S}(t)\ket{S}\bra{S} +
	E_{T_+}\ket{T_+}\bra{T_+} +
	f(t)\left(\ket{S}\bra{T_+} + \mm{H.c.}\right),
	\label{eq:eff_ham}
\end{equation}
where $E_{S}$ is the unperturbed singlet energy, $E_{T_+} = g^* \mu_{\mm{B}} (B
+ B_{\mm{HF},1}^z + B_{\mm{HF},2}^z)$ is the triplet energy, with $g^* = -0.44$ the
effective Landé $g$-factor, $\mu_{\mm{B}}$ the Bohr magneton, $B$ the external magnetic
field, and $B_{\mm{HF},j}^z$ the $z$-component of the hyperfine field in dot $j=1,2$. The
effective coupling $f(t)$ between electronic spin states depends on the hyperfine
interaction with nuclear spins and on the charge state. It can be written as $f(t) = c(t)
\lambda$, with $c(t)$ the time-dependent $(1,1)$ charge amplitude and $\lambda$ the
hyperfine matrix element between $S(1,1)$ and $T_+$. The Lindblad operators
$L_i$ are given by $L_1 = \sqrt{\Gamma_{+}} \sigma_+$, $L_2 = \sqrt{\Gamma_{-}}
\sigma_-$, and $L_3 = \sqrt{\Gamma_{\varphi}} \sigma_z$. They respectively describe
relaxation from excited to ground state and vice versa with rates $\Gamma_{-} = \gamma_1
(n+1)$ and $\Gamma_{+} = \gamma_1 n$ due to phonon emission and absorption, with the mean
phonon number $n=(\ue^{ \Delta E/  k_{\mm{B}}T}-1)^{-1}$ and the spontaneous spin relaxation
rate  $\gamma_1 = 1/T_1$, as well as pure dephasing with a rate $\Gamma_{\varphi}$.
A phenomenological model for the rates leads to the relation $\Gamma_{+} + \Gamma_{-} =
\gamma_1 \coth(\Delta E(t)/ 2 k_{\mm{B}}T)$, where $\Delta E(t)$ is the energy difference
between the instantaneous eigenstates of Eq.~\eqref{eq:eff_ham}, $k_{\mm{B}}$ is
Boltzmann's constant, and $T$ is the phonon bath temperature ($\sim 10\,\mm{mK}$).

We furthermore assume that pure dephasing is mainly due to charge noise when the qubit is
in a superposition of $S(2,0)$ and $T_+$. Since these two states have different
orbital wave functions, they are sensitive to electric fluctuations of the charge
background~\cite{coish2005,hu2006}. We thus assume $\Gamma_{\varphi} = \gamma_2
(1-\abs{c(t)}^2)$. The rates $\gamma_1$ and $\gamma_2$ are free parameters and can be
used to fit experimental results. Nuclear spin induced dynamics are obtained by averaging
solutions of the master equation over a Gaussian distribution of hyperfine
fields~\cite{khaetskii2002,coish2005}, suitable when the thermal energy is larger than
nuclear Zeeman energy, $k_{\mm{B}} T \gg g_{\mm{n}}\mu_{\mm{n}}B$, where $g_{\mm{n}}$ is
the nuclear g-factor and $\mu_{\mm{n}}$ is the nuclear magneton. This description of the
nuclear state is only valid when its internal dynamics happens on characteristic time
scales longer than those of the LZSM driven system (classical approximation). The
standard deviation of the distribution of nuclear fields $B_{\mm{HF},j}^i$
($i=\{x,y,z\}$, $j=\{1,2\}$) is denoted by $\delta_j^i$. The singlet energy and charge
amplitude coefficient used for our simulations are determined
experimentally~\cite{petta2010}.

We consider two types of pulses to measure the singlet return probability
$P_{S}$~\cite{petta2010}. Convolved pulses which are obtained by convolving a
trapezoid pulse with a finite rise-time of $1.5\,\mm{ns}$, a maximal amplitude of
$-2\,\mm{mV}$ and a variable width $t_{\mm{w}}$, with a Gaussian pulse of mean $\mu=0$
and standard deviation $s=3.7\,\mm{ns}$ [see Fig.~\ref{fig:expresults}(a)]. ``Double
hat'' pulses are tailored to have a detuning-dependent level velocity at the leading and
trailing edges of the pulse. The leading edge of the pulse has a level velocity that
varies in the sequence fast-slow-fast. The leading edge has a rise time of $0.1\,\mm{ns}$
and an amplitude of $-2\,\mm{mV}$, which is followed by a slow ramp with a rise time
$t_{\mm{slow}}=8\,\mm{ns}$ and amplitude of $-0.5\,\mm{mV}$. A $0.1\,\mm{ns}$ rise-time
pulse shifts the detuning to its maximal value of $-3\,\mm{mV}$, where the detuning is
held constant for a time interval $t_{\mm{w}}$. The lever-arm conversion between gate
voltage and energy is $\sim$0.13 meV/mV. The trailing edge of the pulse is simply
the reverse of the leading edge [see Fig.~\ref{fig:expresults}(b)]. We present in
Fig.~\ref{fig:expresults}(c) and (e) $P_{S}$ as a function of final detuning
$\varepsilon_{\mm{s}}$ and waiting time $t_{\mm{w}}$ obtained respectively with convolved
pulses for $B=50\,\mm{mT}$ and ``double hat'' pulses for $B=55\,\mm{mT}$.

Since $P_{S}$ for convolved pulses exhibits features already discussed
in~\cite{petta2010}, we only discuss the interference pattern obtained with
``double hat'' pulses. Since the maximal amplitude of these pulses does not depend on
$t_{\mm{w}}$, we can observe interference fringes that start at $t_{\mm{w}}=0\,\mm{ns}$,
which is a first step for manipulation within $T_2^*$. More importantly, we notice three
distinct regions for detunings smaller than $\varepsilon_{\mm{s}}\sim -1\,\mm{mV}$,
which correspond to different magnitude ranges for $P_{S}$. There is an alternation
between regions with $P_{\mm{S}}\simeq1$, $P_{\mm{S}}\simeq 0.4\sim0.9$, and again
$P_{\mm{S}}\simeq1$ in correspondence with the different level velocities associated to
the ``double hat'' pulse. A passage through the anticrossing with a slower level
velocity improves the oscillation visibility, as we could expect from the earlier
considerations within finite-time LZSM theory.

To demonstrate that a high oscillation visibility can be achieved with ``double hat''
pulses, we compare two different types of traces. First, we compare traces taken for a
fixed waiting time. This is equivalent to measuring the visibility of Stückelberg
oscillations for a double passage as a function of $\varepsilon_{\mm{s}}$.  The results
are presented in Figs.~\ref{fig:expresults}(g) and (h) for convolved pulses and ``double
hat'' pulses. To quantitatively compare the visibility of the coherent oscillations we have to neglect the first interference fringe,
which corresponds to a final detuning located at the position of the anticrossing, $\varepsilon_{\mm{s}}=\varepsilon_{\mm{c}}$, which is strongly affected by relaxation mechanisms (cf.\ Refs. \cite{petta2010,ribeiro2011}). We thus find that the
visibility for convolved pulses is $\sim$0.17 and the visibility of ``double hat''
pulses is $\sim$0.5, which corresponds to an improvement of a factor $\sim$2.9. Second,
we present a comparison of traces taken at a fixed value of detuning. This is equivalent
to measuring the visibility of Rabi oscillations. The results are presented in
Figs.~\ref{fig:expresults}(d) and (f) for convolved pulses and ``double hat'' pulses. Neglecting once more the first oscillation dip, we find, by considering only the first peak and relevant
dip, for convolved pulses a visibility of $\sim$0.14 and for ``double hat'' pulses a visibility of $\sim$0.4. Here, there is
an improvement of a factor of $\sim$2.9, which is obtained with $\Delta t \ll T_2^*$. By
considering the first three peaks and dips, i.e. $\Delta t \sim T_2^*$, we find an
improvement of $\sim$2.4. The reduction of visibility is due to nuclear spin dephasing.
We expect to obtain improvements in the visibility close to $\sim$2.9 for suitably
prepared nuclear states \cite{foletti2009}, which exhibit longer decoherence times. The error on the
visibility is on the order of the error on $P_{S}$, which we find to be $\sim$7$\%$.

\begin{figure}
\includegraphics[width=1\columnwidth]{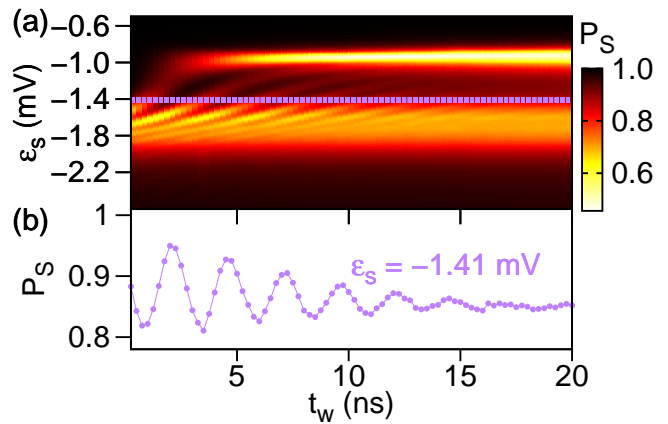}
\caption{(color online) (a) Theoretical calculations of $P_{S}$ as a function of
$t_{\mm{w}}$ and $\varepsilon_{\mm{s}}$ for ``double hat'' pulses and $B=55\,\mm{mT}$.
Theory is in qualitative agreement with the experimental measurements. (b) Trace taken along
$\varepsilon_{\mm{s}}=-1.41\,\mm{mV}$.}
\label{fig:thresults}
\end{figure}

To support our experimental findings, we present in Fig.~\ref{fig:thresults} theory
results obtained by using the experimental pulse profiles measured at the output port of
the waveform generator. We use $\delta^{x,y,z}_{1,2} = 1.00\,\mm{mT}$, $\gamma_1 =
10^5\,\mm{s}^{-1}$, and $\gamma_2 = 10^8\,\mm{s}^{-1}$. Moreover, since the experimental data are acquired at a high rate with cycles of 5 $\mu$s length, we can observe a build up of nuclear polarization. To take this into account in our model, we allow a
nonzero mean for $B_{\mm{HF},j}^i$. The mean $\xi^z_{1,2} \simeq 0.0\,\mm{mT}$ for
$B_{\mm{HF},1,2}^z$ can be determined from spin-funnel measurements~\cite{petta2010}.
Since we cannot experimentally determine $\xi_{1,2}^{x,y}$, we have chosen $\xi_1^{x,y} =
6\,\mm{mT}$ and $\xi_2^{x,y} = 0\,\mm{mT}$. Our theory results agree qualitatively with
the experiments, as can be seen when comparing interference fringes [see
Figs.~\ref{fig:thresults}(a) and \ref{fig:expresults}(e)].

Our results indicate that the qubit is not only influenced by nuclear spins, but that
there are additional physical mechanisms that determine the oscillation visibility. Here,
the contrast is also limited due to the superposition of $S(2,0)$ and
$S(1,1)$~\cite{ribeiro2011}. First, the weighting of $S(1,1)$ sets the amount
of population that can be transferred to $T_+$. Second, superpositions of different
charge states are susceptible to charge noise, which results in an additional effective
spin dephasing mechanism. This dephasing channel directly competes against LZSM tunneling
by preventing the qubit from coherently interfering with itself. Spin relaxation also
changes the balance of populations, but due to energy scales its effect is weak far from
the avoided crossing, where $k_{\mm{B}} T \ll \Delta E$.

In conclusion, we have demonstrated how to increase the visibility of quantum
oscillations by enhancing the adiabatic passage probability in the presence of
dissipation. We have designed a pulse which combines both fast and slow rise-time ramps
to minimize dissipation and enhance adiabaticity. By considering a $S$-$T_+$
anticrossing, we have shown that it is possible to achieve coherent superposition states
with high $T_+$ population. In the more general context of LZSM driven spin qubits,
this technique allows one to perform more quantum gates within a given decoherence time
and achieve higher amplitude rotations in the qubit space without exponentially extending
the gate operation times. Our control technique can be further improved by preparing a
nuclear spin gradient~\cite{foletti2009}. This will not only increase $T_2^{*}$, but it
will also enhance the effective coupling between spin states, thus boosting adiabatic
transition probabilities.

\begin{acknowledgments}
We acknowledge fruitful discussions with David Huse and Mark Rudner. Research at
Princeton is supported by the Sloan and Packard Foundations, the NSF through the
Princeton Center for Complex Materials (DMR-0819860) and CAREER award (DMR-0846341), and
DARPA QuEST (HR0011-09-1-0007). Work at UCSB was supported by DARPA (N66001-09-1-2020)
and the UCSB NSF DMR MRSEC. H. R. and G. B. acknowledge funding from the DFG within SPP
1285 and SFB 767.
\end{acknowledgments}

\end{document}